\documentclass[12pt]{article} 

\usepackage{publication.oneside} 

\usepackage[T1]{fontenc}
\usepackage[english]{babel}
\usepackage[utf8]{inputenc} 
\usepackage{graphicx}
\usepackage{amssymb}
\usepackage{amsfonts}
\usepackage{amsmath}
\usepackage{mathtools}
\usepackage{cite} 

\usepackage[footnotesize]{caption} 
\usepackage[font=footnotesize,margin=10pt,format=hang]{subfig}

\def\iA{\mathcal{A}}
\def\iH{\mathcal{H}}
\def\iE{\mathcal{E}}
\def\Tr{\trace\,}
\def\<{\langle}
\def\>{\rangle}
\def\bbbc{{\mathbb C}}

\begin{document}
\begin{center}
\Large{\sc Experiment design and parameter estimation of Pauli channels using convex optimization}
\vspace{0.5cm}\\ 
\textbf{\normalsize{Gábor Balló}}\\
\small{Department of Electrical Engineering and Information Systems,\\University of Pannonia,\\H-8201 Veszpr\'{e}m, Egyetem u. 11, Hungary,\\e-mail: \textit{ballo.g@virt.uni-pannon.hu}}
\vspace{0.5cm}\\ 
\textbf{\normalsize{Katalin M. Hangos}}\\
\small{Process Control Research Group of the Computer and Automation Research Institute,\\H-1111 Budapest, Kende u. 13-17, Hungary,\\e-mail: \textit{hangos@scl.sztaki.hu}}
\vspace{0.5cm}\\ 
\large{\today}
\thispagestyle{empty} 
\end{center}
\vspace{1cm}

\begin{abstract}
A unified framework is proposed in this paper for parameter estimation using 
convex optimization and experiment design applying convex maximization 
 for Pauli channels, that can be extended to generalized Pauli channels, too. 
In the case of known channel directions, an affine parametrization of the Choi matrix turns 
 the LS parameter estimation into a convex optimization problem also for the generalized Pauli channels. 
A simple iterative algorithm for estimating the channel directions is also given for qubit Pauli channels. 

The experiment design was performed by maximizing the trace of the Fisher information 
 matrix of the output 
 quantum system to find optimal input state and measurement POVM for the channel estimation. 
For the known channel direction case it was found that the optimal input state should be pure and 
 the optimal measurement POVM is extremal. 
It was also shown that both the input state and the POVM elements in the optimal configuration 
 should be parallel to the channel directions in the qubit Pauli channel case. 

The proposed methods and algorithms are illustrated by simple numerical examples. 
\end{abstract} 


\newpage
\section{Introduction}

Quantum systems are special stochastic nonlinear systems, where the stochasticity and nonlinearity are caused by the back-action of the measurements on the measured system \cite{Nielsen-book}, \cite{Petz-book}. 
Therefore, even in the simplest static case, when the parameters of a non-dynamic quantum system are to be estimated, one needs special estimation methods \cite{ParReh:2004}; this case is called state tomography in theoretical quantum physics. 
  
Quantum channels are widely used information transfer devices in quantum information theory \cite{Nielsen-book}, that map an input quantum system into an output one usually in a static way. 
The task of the estimation of quantum channels -- commonly known as \textit{quantum process tomography} (QPT) in theoretical quantum physics  -- got a significant attention over about the last ten years.  
The problem was investigated by several authors \cite{Nielsen-book}, \cite{Dariano-2003}, \cite{Paris-book}. The work \cite{Mohseni-2008} gives a comprehensive survey on the different strategies used for process tomography (or channel estimation). The problem can essentially be formulated in two type of methods: direct, and indirect. In the indirect method, we trace the problem back to quantum state tomography, i.e.\ the information about the unknown quantum channel is obtained by sending known probe quantum systems through the channel, and performing state tomography on the output states. In contrast, in the direct method, the experiments directly give information about the channel, without the need for a state tomography step.

From a methodological point of view, there are two principally different approaches to the problem of quantum tomography, the \emph{statistical approach} and the convex optimization based approach \cite{Dariano-2003}. The former gives information on the statistics of the estimate and on its covariance matrix, but it has the drawback, that it is hard to compute in higher dimensions. In spite of this, majority of the existing methods belong to this category. 

In contrast to this, an \emph{optimization based method} does not give as much information, but it is relatively easy to compute. This approach has been pursued in the work \cite{Sacchi-2001} where the problem of channel estimation (in the form of the Choi matrix, which is a Hermitian matrix representing the channel) is considered, assuming a completely general channel, thus without any assumption on the inner structure of the Choi matrix. The author uses random input states, and random measurements on the output, and formulates a maximum likelihood problem. A similar method is used in the work \cite{Kosut-2004}, which formulates the task of process tomography as a least squares problem, which is convex. It also uses the Choi matrix as optimization variable, thus searches the optimal channel in the convex set of all \emph{completely positive and trace preserving} (CPTP) maps using multiple input-measurement pairs. 

However, as it is stated in \cite{Sasaki-2002}, it is a reasonable assumption to consider only a certain family of channels given with a model, based on a priori knowledge about the structure of the channel. One of the aims of our work is to develop a method which is capable of incorporating these constraints into the channel tomography problem, while still remaining -- at least partially in the general case -- solvable by convex optimization. 

It is a commonly known fact that system identification is intimately related to \textit{experiment design}, the general aim of which is determining experimental conditions that result in good or even optimal identification results \cite{Ljung1999}. Thus the method of identification, that is, model parameter and structure estimation, determines the methods applied for experiment design, too. In addition, the nature and properties of the system to be identified have also a determining influence on identification and experiment design. The experiment design for quantum channel parameter estimation includes the design of the quantum input to the channel, and the observables to be applied on the resulting quantum output system, that is called the \textit{experiment configuration}, together with the number of measurements to be performed in the different experiment configurations if one has a few of them. 

The results on experiment design for quantum state and channel estimation 
 appear sparsely in the quantum state and process tomography literature, 
when the authors investigate the optimality of their experiment configurations. 
The problem of optimal \textit{experiment design for quantum state tomography} was first 
 investigated by Kosut et al.\  \cite{Kosut-2004} who developed methods using 
 convex optimization for the 
determination of the number of measurements to be performed in the different experiment configurations. 
Since then, a few more papers can be found about optimal experiment design for quantum state 
 estimation (see e.g.\ \cite{Nunn2010} for a recent paper), but the problem is far from being solved for all cases. 
 
The problem of finding an \textit{optimal estimation of} one parameter \textit{quantum channels} is discussed in \cite{Sarovar2006}, 
 for both the single qubit input and the entangled qubit input case using statistical methods. 
An efficient estimation scheme is proposed in \cite{Fujiwara2003}, where the quantum Fisher 
 information and information geometrical considerations lead to an optimal measurement configuration 
for generalized Pauli channels that act on more than 2-dimensional quantum systems 
 with entangled 
 finite dimensional quantum inputs. 
Although the restricted experiment design problem, i.e.\ the 
 determination of the number of measurements to be performed in the different experiment configurations 
 for quantum channels has also been formulated in the pioneering work of \cite{Kosut-2004}, it is much less investigated 
 than its state tomography counterpart. 
A recent paper of \cite{Branderhorst2009} gives a good overview of the state-of-the-art in this field. 
 
Motivated by the above experiment design problems for quantum process tomography and by our 
 recent work of optimization based quantum channel estimation \cite{Ballo-2010}, the aim of this study 
is to propose a unified framework for parameter estimation using convex optimization
 and experiment design applying convex maximization 
 for Pauli channels, that can be extended to generalized Pauli channels, too. 

The rest of this paper is organized as follows. 
First the basic notions on quantum systems, quantum channels and quantum parameter 
 estimation are described for the finite dimensional case in the next section. 
Then the convex optimization-based channel parameter estimation method is 
 given in section \ref{sec:qpt.as.opt.problem}. 
This is followed by a section about estimating the channel directions. 
Section \ref{sec:exp.des} is devoted to the proposed experiment design 
 method that is based on convex maximization. 
A separate section thereafter illustrates the advantages of the proposed experiment 
 design method using case studies. 
Finally conclusions are drawn. 

\section{Basic notions}

Some basic notions of quantum systems are described in this section for the finite dimensional case.  

\subsection{Quantum Measurements and Fisher Information}

\subsubsection{State Representation of Finite Dimensional Quantum Systems}

The state of a finite dimensional quantum system is described 
by a so called density operator or \textit{density matrix} $\rho$ that acts 
on the underlying finite dimensional complex Hilbert space $\mathcal{H}$. 
Density matrices are self-adjoint positive semidefinite matrices with unit trace, i.e.\
\begin{equation}
\rho \ge 0, \quad \rho^{*}=\rho, \quad \trace(\rho)=1\ .
\end{equation}
where $\rho^*$ denotes the adjoint of $\rho$. 

Two-level quantum systems are called \textit{quantum bits}, their density 
 matrices are $2 \times 2$ complex matrices that are of the form
\begin{equation}\label{E:2x2}
\rho= \frac{1}{2}\left[\begin{array}{cc}
1+\theta_3 & \theta_1 - i \theta_2 \\
\theta_1 + i \theta_2 & 1-\theta_3 \end{array}\right]=
\frac{1}{2}\left(I + \sum_{i=1}^3 \theta_i \sigma_i \right)\ ,
\end{equation}
where $\sigma_1, \sigma_2,\sigma_3$ are the so-called Pauli matrices, and 
$I$ is the unit matrix. 
The vector $\theta=(\theta_1,\theta_2,\theta_3)^\tp$ is in the 3-dimensional 
unit ball of $\mathbb{R}^{3}$. 
This state representation is called \textit{Bloch vector}. 

Let $|\phi_{i,1}\>$ and $|\phi_{i,2}\>$ be the normalized eigenvectors of 
$\sigma_i$ ($i=1,2,3$). Then
$$
\sigma_i= |\phi_{i,1}\>\<\phi_{i,1}|-|\phi_{i,2}\>\<\phi_{i,2}| \qquad (i=1,2,3).
$$
It is well-known that
\begin{equation}\label{E:mub}
| \< \phi_{i,k} |\phi_{j,l}\>|^2=\frac{1}{2} \quad \quad (i\neq j)
\end{equation}
which means that the three bases $\{ \ket{\phi_{i,1}}, \ket{\phi_{i,2}} \} $  ($i=1,2,3$)
are \emph{mutually unbiased}, and form a so called \textit{MUB} (mutually unbiased bases).

In this formalism we have
\begin{eqnarray}\label{eq:mubpaulichannel}
\rho& = &\frac{I}{2}+ \frac{1}{2}\sum_{i=1}^3 
\theta_i \left( |\phi_{i,1}\>\<\phi_{i,1}|-|\phi_{i,2}\>\<\phi_{i,2}| \right) \cr
&=& \frac{1}{2}\left(1- \sum_{i=1}^3 \theta_i\right)I+\sum_{i=1}^3 
\theta_i |\phi_{i,1}\>\<\phi_{i,1}|\ .
\end{eqnarray}
Note that we can uniquely represent any density matrix with its Bloch vector $\mathbf{\theta}$ given the MUB. 
 
\subsubsection{Quantum Measurements}

Quantum measurements are described mathematically as  \\ 
$\n{M}=\{M_1,...,M_m \}$, where
the  self-adjoint positive operators $M_i$ act on the Hilbert space and
$M_1+...+M_m=I$. Such $\n{M}=\{M_1,...,M_m \}$ is 
called to be a \emph{positive operator-valued measure} (POVM). If the positive operators $M_i$ are all projections, then we get a so called projective measurement. 

If a POVM $\n{M}$ is performed
as a measurement on the state $\rho$, then the possible outcomes are $1,2,\dots, m$
and the probability of the outcome $i$ is $\trace(\rho M_i)$. Let $\rho_\theta$ be a parametrized quantum state and a POVM $\n{M}=\{M_\alpha: \alpha 
\in A\}$ is used for the measurement, where $A$ denotes the set of measurement outcomes. 
Thus the probability distribution of the outcomes is
\begin{equation} \label{eq:prob.distribution.of.measurement}
p(\alpha|\theta)=\trace(\rho_\theta M_\alpha) \qquad (\alpha \in A)\ ,
\end{equation}
which is a set of probability distributions parametrized by $\theta$.
 
\subsubsection{Fisher Information} \label{sec:fisherdef}

The Fisher information reflects the amount of information that a measured random variable 
can carry about the parameter $\theta=(\theta_1,\dots, \theta_k)^\tp$. In other words, it 
measures the accuracy of the unbiased estimator $\hat\theta$ of $\theta$.  Fisher 
information is a classical concept in statistics.

The accuracy of the estimator is expressed by the covariance matrix:
\[
\mathrm{Var}(\hat\theta)_{i,j}= E(\hat\theta_i \hat\theta_j)- E(\hat\theta_i)E( \hat\theta_j)\ ,
\]  
where $E$ is the expectation value.

The Fisher information matrix for the quantum setting is by \cite{Gill-2002}
\begin{equation} \label{Eq:fisherdef}
F(\theta)_{i,j}=\sum_\alpha \frac{1}{\trace(\rho_\theta M_\alpha)}
\frac{\partial}{\partial \theta_i}\trace(\rho_\theta M_\alpha)
\frac{\partial}{\partial \theta_j}\trace (\rho_\theta M_\alpha)\ ,
\end{equation}
and the Cram\'er--Rao matrix inequality describes their relation: 
\[
\mathrm{Var}(\hat\theta)\geq F(\theta)^{-1}
\]
This bound shows that the higher the Fisher information, the better estimation we can have. 
The formula for $F$ also shows that in the quantum case, it depends on the actual
measurement POVM $\n{M}$ with which the experiments had been performed, i.e.\ 
$F(\theta,\n{M})$. 

\subsection{Quantum Channels}

Quantum channels model the information transfer between quantum systems, i.e.\ they transform 
 the source quantum system into a target one. A quantum channel
 $\cha{E}: \mathcal{B}(\spa{H}_1) \to \mathcal{B}(\spa{H}_2)$  is defined to be a completely positive 
 and trace preserving (CPTP) map, where $\mathcal{B}(\spa{H}_i)$ is the operator algebra on the Hilbert space $\spa{H}_i$. 
This map can be represented by a set of operators, $V_i:\spa{H}_1 \to \spa{H}_2$,
called the Kraus representation, which gives the channel output as
\begin{equation}
\cha{E}(\rho)=\sum_i V_i\rho V_i^*\ ,
\label{eq:kraus.form}
\end{equation}
and the operators must satisfy the relation 
\begin{equation}
\sum_i V_i^*V_i=I\ ,
\label{eq:tp.constraint}
\end{equation}
in order to represent a trace preserving map. 

We have to mention that the set of operators in the above representation is not unique. 
This drawback can be eliminated with another possible description of channels using the 
definition of the \textit{Choi matrix}. Let $\spa{H}_1$ and $\spa{H}_2$ be Hilbert spaces, and 
$\mathcal{E}: \mathcal{B}(\mathcal{H}_1) \to \mathcal{B}(\mathcal{H}_2)$ be a linear 
 mapping that represents the quantum channel. 
To define the Choi matrix of $\mathcal{E}$ we take an orthonormal basis $f_1,...,f_n$ in 
$\mathcal{H}_1$. Then $\ket{f_i}\bra{f_j} \in \mathcal{B}(\mathcal{H}_1)$ and $\mathcal{E}$ acts on this 
operator. Then the Choi matrix according to \cite{Petz-book} is
\begin{equation} \label{eq:Choidef}
 X_{\mathcal{E}} = \sum_{i,j}  \ket{f_i}\bra{f_j} \otimes  \mathcal{E} \big(\ket{f_i}\bra{f_j} \big)
\in \mathcal{B}(\mathcal{H}_1)\otimes \mathcal{B}(\mathcal{H}_2)\ .
\end{equation}
where $\otimes$ denotes the tensor product. 
Actually, the above matrix is a block matrix, its $ij$th element is $\mathcal{E}\big( \ket{f_i}\bra{f_j} \big)$. 
The complete positivity of $\mathcal{E}$ is equivalent with the positivity of $X_{\mathcal{E}}$. 
Furthermore, $\mathcal{E}$ is trace preserving, if 
$$
\trace\left[\mathcal{E}\big( \ket{f_i}\bra{f_j}\big)\right] = \trace\big(\ket{f_i}\bra{f_j}\big) = \delta_{ij}
$$ 
which means $\trace_2(X_{\mathcal{E}})=I$.

\subsection{Pauli Channels}
A notable wide class of quantum channels are the \emph{Pauli channels}. 

\subsubsection{Qubit Pauli Channels}
\label{ssec:qpau}
In the qubit case when the input density matrix $\rho$ has the form (\ref{E:2x2}),  
\begin{equation}
\label{eq:2x2out}
\mathcal{E}(\rho)=\frac{1}{2}\left[\begin{array}{cc}
1+\lambda_3 \theta_3 & \lambda_1\theta_1 - i \lambda_2\theta_2 \\
\lambda_1 \theta_1 + i \lambda_2\theta_2 & 1-\lambda_3 \theta_3 \end{array}\right]=
\frac{1}{2}\left(I + \sum_{i=1}^3 \lambda_i \theta_i \sigma_i \right).
\end{equation}
is a simple example of a qubit Pauli channel. The Choi matrix of this channel 
is the following:
\[
\op{X}_{\cha{E}}=\frac{1}{2}\left[ \begin {array}{cccc}  1+\lambda_3 & 0& 0&
    \lambda_1+\lambda_2\\\noalign{\medskip} 0&
    1-\lambda_3 & \lambda_1-\lambda_2&
    0\\\noalign{\medskip} 0& 
\lambda_1-\lambda_2 & 1-\lambda_3 & 0\\\noalign{\medskip}
\lambda_1+\lambda_2 & 0& 0& 1+\lambda_3 \end {array} \right]\ .
\]
The conditions of positivity for this matrix in terms of the parameters are
\begin{equation} \label{eq:trace.preserving}
|1\pm\lambda_3|\geq|\lambda_1\pm\lambda_2|\ ,
\end{equation}
and the trace preserving requires 
\begin{equation} \label{eq:tpres2}
|\lambda_i|\leq 1~~,~~i=1,2,3 . 
\end{equation}

The typical Pauli channel acts on a density matrix
$$
\rho=\frac{1}{2}\left(1- \sum_{i=1}^3 \theta_i\right)I+\sum_{i=1}^3 
\theta_i |\phi_{i,1}\>\<\phi_{i,1}|
$$
as
\begin{equation} \label{eq:2pauli}
\iE(\rho)=\frac{1}{2}\left(1-\sum_{i=1}^3 \lambda_i \theta_i\right)I+\sum_{i=1}^3 
\lambda_i \theta_i |\phi_{i,1}\>\<\phi_{i,1}|.
\end{equation}
To describe this channel we need the three real constants $\lambda_1, \lambda_2, \lambda_3$
(satisfying (\ref{eq:trace.preserving}) and (\ref{eq:tpres2})) and the vectors 
$|\phi_{1,1}\>, |\phi_{2,1}\>,|\phi_{3,1}\>$ (satisfying (\ref{E:mub})). 
So a Pauli channel is given by 6 data items.
Below the vectors $|\phi_{1,1}\>, |\phi_{2,1}\>,|\phi_{3,1}\>$ will be called \emph{channel directions}. The 
$$
\iA_i=\{ a |\phi_{i,1}\>\<\phi_{i,1}| + b |\phi_{i,2}\>\<\phi_{i,2}| : a,b \in
\bbbc\} \qquad (i=1,2,3)
$$
are commutative subalgebras. The effect of the channel can then be described as depolarizing in each subalgebra $\iA_i$ with the corresponding parameter $\lambda_i$.

\subsubsection{Generalized Pauli Channels}
\label{sec:gpauli}

The generalized Pauli channel is discussed in \cite{Petz-2008}. Assume that
the operator algebra $\spa{B}(\spa{H})$ contains subalgebras $\iA_1,\iA_2, \dots , \iA_u$ 
which are complementary in the sense that
$$
\Tr (A_i A_j)=0 \mbox{ if } \Tr(A_i)=\Tr (A_j)=0, A_i \in \iA_i, A_j \in \iA_j, i
\ne j.
$$
The trace preserving projection $E_i:\spa{B}(\spa{H}) \to \iA_i$ is usually
called conditional expectation. If $d$ is the dimension of $\spa{H}$, then for 
 any input $A\in\mathcal{B}(\mathcal{H})$ in the form 
$$
 \op{A}=- \frac{(u-1) \Tr(\op{A})}{d} + \sum_{i=1}^{u} E_i (\op{A})~,
$$
\begin{equation}\label{E:gpc}
\cha{E}(\op{A})=\left(1-\sum_{i=1}^{u}\lambda_i\right)
\frac{\trace(\op{A})}{d}I +\sum_{i=1}^{u}\lambda_iE_i(\op{A})
\end{equation}
is a generalization of the Pauli channel. In the classical Pauli setting the
$$
\iA_i=\{ a |\phi_{i,1}\>\<\phi_{i,1}| + b |\phi_{i,2}\>\<\phi_{i,2}| : a,b \in
\bbbc\} \qquad (i=1,2,3)
$$
are commutative subalgebras and, for example 
$$
E_1 \left(\left[\begin{array}{cc}
1+ \theta_3 & \theta_1 - i \theta_2 \\
\theta_1 + i \theta_2 & 1-\theta_3 \end{array}\right]\right)=
\left[\begin{array}{cc}
1 & \theta_1  \\ \theta_1   & 1 \end{array}\right].
$$
Similarly to this example here we consider the case when all of the
complementary subalgebras are maximal Abelian. 

In this case, we can take $u$ orthonormal bases in the $d$ dimensional Hilbert space $\iH$:
$$
\{ |\phi_{i,k}\>: 1 \le k \le d\} \qquad 1 \le i \le u\ .
$$
Similarly to the Pauli case, it is assumed that
\begin{equation}
|\braket{\phi_{i,k}}{\phi_{j,l}}|^2=\frac{1}{d} \quad \quad (i\neq j)\ ,
\end{equation}
i.e. the bases are mutually unbiased.
Then the relevant subalgebras are
$$
\iA_i=\left\{ \sum_{k=1}^d c_k |\phi_{i,k}\>\<\phi_{i,k}|:c_k \in \bbbc\right\} \qquad
(1 \le i \le u)\ ,
$$
and the conditional expectations are 
$$
E_i(A)=\sum_{k=1}^d \<\phi_{i,k}, A \phi_{i,k}\> |\phi_{i,k}\>\<\phi_{i,k}|\ .
$$
So the generalized Pauli channel is defined by the formula (\ref{E:gpc}). The
conditions for complete positivity and trace preserving property are 
\begin{equation} \label{eq:genPauli_inequ}
1+d \lambda_i \ge \sum_j \lambda_j \ge - \frac{1}{d-1}\ ~~~,~~~| \lambda_i| \leq 1 .
\end{equation}

\section{Quantum Channel Parameter Estimation as an Optimization Problem}
\label{sec:qpt.as.opt.problem}
The parameter estimation of quantum channels, or quantum process tomography is a widely investigated 
 problem in mathematical physics. 
The pioneering works of Kosut and co-workers \cite{Kosut-2004} formulated its variants as a 
convex optimization problem. 

The formal mathematical description of the general quantum process tomography problem is as follows. 
Consider an unknown quantum channel $\cha{E}: \spa{B}(\spa{H})\rightarrow\spa{B}(\spa{H})$, which is to be estimated. We use a so called \emph{tomography configuration} for this purpose that contains the following elements: 
\begin{itemize}
\item A known input density operator $\rho$ on the Hilbert space $\spa{H}$ of the system. 
\item A POVM with which we can perform quantum measurement on the channel output state $\cha{E}(\rho)$. 
\end{itemize}
From the above, we compute an estimate of the channel parameters using the measured data and the parametrized 
 model of the channel. 

Note that we can use multiple different tomography configurations, i.e. different input states and POVMs in order to achieve better estimation on $\cha{E}$. In this work, the (input, POVM) pair corresponding to the $\gamma$th configuration is denoted by $(\rho_\gamma,\textbf{M}_\gamma)$. 

\subsection{The Estimation Method}
\subsubsection{Experimental Data Collection}
\begin{figure}
\im[width=\columnwidth]{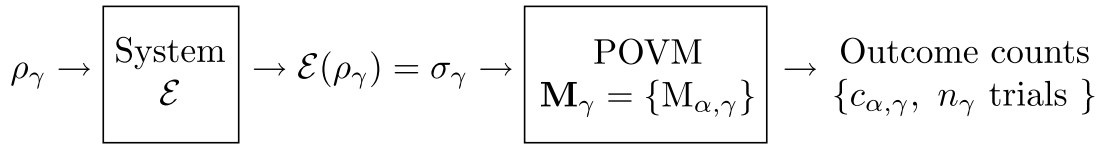}
\caption{The scheme of data collection for process tomography.\label{fig:data.collection}}
\end{figure}

The first stage of process tomography is the collection of the measurement data. The measurements are performed in each $\gamma$ configuration $n_\gamma$ times independently. During data collection, the different $\alpha$ outcomes of the measurements in the configuration $\gamma$ are counted in the variable $c_{\alpha,\gamma}$, and put in the measurement record $\textbf{R}$. 
This scheme can be seen in Fig. \ref{fig:data.collection}.
Then obviously $\sum_\alpha c_{\alpha,\gamma} = n_\gamma$. Thus, we have to perform a total number of $n_\tn{tot}=\sum_\gamma n_\gamma$ independent measurements. The estimator $\hat{\cha{E}}$ of the channel $\cha{E}$ will be calculated from these measurement outcomes.

\subsubsection{Least Squares Estimation}
If $\cha{E}$ were given in its Kraus representation, then the resulting LS optimization problem would be nonconvex, because the optimization variables would be the Kraus operator elements $\op{V}_i$. To overcome this difficulty, it is reasonable to choose the Choi matrix as optimization variable. 

By the use of relations (\ref{eq:prob.distribution.of.measurement}) and (\ref{eq:Choidef}) we get \cite{Ballo-2010}
\[
p_{\alpha,\gamma}= \trace(C_{\alpha,\gamma}X_{\cha{E}})\ ,
\]
where $\op{X}_{\cha{E}}$ is the Choi matrix of the channel $\cha{E}$, and the configuration matrix $C_{\alpha,\gamma}=\rho_\gamma^\tp\otimes\op{M}_{\alpha,\gamma}^*$ depends on the channel input $\rho$ and on the measured POVM elements in configuration $\gamma$. 

The probability $p_{\alpha,\gamma}$ can be estimated as 
\begin{equation}
\label{eq:ls.relative.frequency}
\hat{p}_{\alpha,\gamma}=\frac{c_{\alpha,\gamma}}{n_\gamma}
\end{equation} 
using the relative frequency that can be calculated from the measurement results. 
The variance of this unbiased estimate after $n_\gamma$ independent measurements is known to be 
\begin{equation} \label{eq:var_prob}
\mathrm{Var}\big(\hat{p}_{\alpha,\gamma}\big)=\frac{1}{n_\gamma}p_{\alpha,\gamma}\big(1-p_{\alpha,\gamma}\big)\ ,
\end{equation}
because $\hat{p}_{\alpha,\gamma}$ has a binomial distribution. These show that for large $n_\gamma$, $\hat{p}_{\alpha,\gamma}\rightarrow p_{\alpha,\gamma}$ and $\mathrm{Var}\big(\hat{p}_{\alpha,\gamma}\big)$ tends to $0$ as $n_\gamma\rightarrow\infty$, so $\hat{p}_{\alpha,\gamma}$ is a reasonable unbiased estimate of the real value $p_{\alpha,\gamma}$. This leads to formulating the parameter estimation as the following least squares problem: 
\begin{gather}
\label{eq:ls.objective}
\arg\min_{\op{X}_{\cha{E}}} \sum_{\alpha,\gamma}\Big[\hat{p}_{\alpha,\gamma}-\trace\big(\op{C}_{\alpha,\gamma}\op{X}_{\cha{E}}\big)\Big]^2\ ,\\
\tn{so that}\quad\op{X}_\cha{E}\geq 0,\quad\trace_2(\op{X}_\cha{E})=I\ \nonumber
\end{gather}

This problem is a convex optimization problem in the Choi matrix $\op{X}_\cha{E}$, thus it can be solved relatively easily using existing numerical algorithms \cite{Vandenberghe-1996,Audenaert-2002}. 

\subsection{Estimation of Pauli Channel Model Families with Known Channel Directions}
\label{sec:est.model.families}

In practice it is reasonable to assume that we know a model type of the channel, and only the unknown values of the parameters of this model have to be estimated. In such a problem, the above derived least squares objective cannot be used directly, as it assumes a completely general channel model, and estimates the elements of the Choi matrix. Thus, if the task is to estimate some specific model parameters, this method can suffer significantly from overparametrization. 

Some authors proposed approaches based on prior information on the channel, thus obtaining a well conditioned parameter estimation problem. These are mainly derived from physical interactions involved in the dynamics \cite{Branderhorst2009}, and in \cite{Young-2009} the authors also consider the problem of finding the optimal series of experiments to estimate the channel parameters.

As another possible solution, we can study the internal structure of the Choi matrix, and use this information to select more appropriate, model specific parameters for optimization. Effectively, this should reduce the set of optimal solutions of problem \eqref{eq:ls.objective} to solutions, which are consistent with the desired model family. 

\subsubsection{Affine Approximation}
\label{ssec:affine}
The natural choice would be to select just the unknown channel parameters. However it can be easily seen, that this choice would ruin convexity, as the Choi matrix can be an arbitrarily nonconvex function of these in the most general case. 

Thus, instead of this, the following method is used. Let $h_1(\lambda),\dots,h_m(\lambda)$ denote functions of the channel parameters, and let $H_0,H_1,\dots,H_m$ denote constant Hermitian matrices. Then we can expand the Choi matrix as an affine function \cite{Ballo-2010}:
\begin{equation}
\label{eq:affine.approx}
\op{X}_\cha{E}=\sum_kH_kh_k(\lambda)+H_0
\end{equation}
This way we can approximate the Choi matrix by an affine structure, and use the functions $h_k(\lambda)$ as optimization variables. 

Note that the trace preserving constraint can be omitted in \eqref{eq:ls.objective}, as it can always be taken into account in the model construction.   

\subsubsection{A Pauli Channel Parameter Estimation Example}

Assume that the channel directions are known or have been determined (see section \ref{sec:direction_est} later).  
Using \eqref{eq:affine.approx}, the Choi matrix of the qubit Pauli channel can be decomposed using the following Hermitian matrices:
\begin{align*}
&H_0=\frac{1}{2}
{\scriptsize 
\left[ 
\begin {array}{cccc} 
 1 &  0  &  0  &  0\\ \noalign{\medskip}
 0 &  1  &  0  &  0\\ \noalign{\medskip}
 0 &  0  &  1  &  0\\ \noalign{\medskip}
 0 &  0  &  0  &  1
\end{array}
\right]
},
&&H_1=\frac{1}{2}
{\scriptsize 
\left[ 
\begin {array}{cccc} 
 0 &  0  &  0  &  1\\ \noalign{\medskip}
 0 &  0  &  1  &  0\\ \noalign{\medskip}
 0 &  1  &  0  &  0\\ \noalign{\medskip}
 1 &  0  &  0  &  0
\end{array}
\right]
},\\
&H_2=\frac{1}{2}
{\scriptsize 
\left[ 
\begin {array}{cccc} 
 0 &  0  &  0  &  1\\ \noalign{\medskip}
 0 &  0  & -1  &  0\\ \noalign{\medskip}
 0 & -1  &  0  &  0\\ \noalign{\medskip}
 1 &  0  &  0  &  0
\end{array}
\right]
},
&&H_3=\frac{1}{2}
{\scriptsize 
\left[ 
\begin {array}{cccc} 
 1 &  0  &  0  &  0\\ \noalign{\medskip}
 0 & -1  &  0  &  0\\ \noalign{\medskip}
 0 &  0  & -1  &  0\\ \noalign{\medskip}
 0 &  0  &  0  &  1
\end{array}
\right]
},
\end{align*}
and the new optimization variables will be
\[
h_1=\lambda_1,\quad h_2=\lambda_2,\quad h_3=\lambda_3\ .
\]
Thus, in this representation, \textit{the parameter estimation of any two dimensional Pauli channel is a convex problem, as the optimization variables are exactly the channel parameters to be estimated}.   

\subsubsection{The Case of Generalized Pauli Channels}

The above example can be generalized to the higher level Pauli channel case 
assuming again known channel directions. 
For the sake of simplicity we choose the simplest 3-level (qutrit) case, when $d=3$ in sub-section \ref{sec:gpauli}. The used MUB will be the bases suggested in \cite{Petz-2007}. Then the Choi matrix of the channel will be the following:
\[
\op{X}_{\cha{E}}= \frac{1}{3}
{\scriptsize 
 \left[ \begin {array}{ccccccccc}  
 f_1  & 0   & 0   & 0   & f_3 & 0   & 0   & 0   & f_3   \\\noalign{\medskip} 
 0    & f_2 & 0   & 0   & 0   & f_4 &f_4^*& 0   & 0     \\\noalign{\medskip} 
 0    & 0   & f_2 &f_4^*& 0   & 0   & 0   & f_4 & 0     \\\noalign{\medskip} 
 0    & 0   & f_4 & f_2 & 0   & 0   &  0  &f_4^*& 0     \\\noalign{\medskip} 
 f_3  & 0   & 0   & 0   & f_1 & 0   & 0   & 0   & f_3   \\\noalign{\medskip} 
 0    &f_4^*& 0   & 0   & 0   & f_2 & f_4 & 0   & 0     \\\noalign{\medskip} 
 0    & f_4 & 0   & 0   & 0   &f_4^*& f_2 & 0   & 0     \\\noalign{\medskip} 
 0    & 0   &f_4^*& f_4 & 0   & 0   & 0   & f_2 & 0     \\\noalign{\medskip} 
 f_3  & 0   & 0   & 0   & f_3 & 0   & 0   & 0   & f_1   \end{array} 
 \right]\ ,}
\] 
where 
\begin{gather*}
f_1=1+ 2\lambda_2,\quad f_2=1-\lambda_2,\quad f_3=\lambda_1+\lambda_3+\lambda_4,\\
f_4=\lambda_1-\frac{\lambda_3}{2}(1+\ii\sqrt{3})-\frac{\lambda_4}{2}(1-\ii\sqrt{3}),
\end{gather*}
Note that the $d=2$ case using the same MUB selection method is exactly the same as the qubit Pauli channel in subsection \ref{ssec:qpau}. 

The decomposition \eqref{eq:affine.approx} of the Choi matrix can be calculated easily given that the optimization variables are again exactly the channel parameters:
\[
h_i=\lambda_i,\quad i=1,\dots,4
\]
This, and the construction of the channel shows that \textit{the problem of parameter estimation is convex, and solvable using only \eqref{eq:ls.objective} in any such -- arbitrarily high -- dimension, in which the channel itself can be defined}. 

\section{Estimating the Channel Directions} 
\label{sec:direction_est}
During the parameter estimation of Pauli channels it is generally assumed that the Pauli channel directions are known. This, however, is not true in general so this section describes a method to estimate these directions, while resulting in a first estimate on the parameters, too \cite{Ballo-2010}. 

\subsection{Qubit Case}
If we do not know the exact three directions $|\phi_{1,1}\>, |\phi_{2,1}\>,|\phi_{3,1}\>$ in which the Pauli channel is depolarizing, then quantum state estimation steps can be used to determine this structure. 

Let us fix three vectors $|\varphi_{1,1}\>, |\varphi_{2,1}\>,|\varphi_{3,1}\>$ satisfying (\ref{E:mub}). 
Then the operators $|\phi_{i,1}\>\<\phi_{i,1}|~,~i=1,2,3$ formed by the channel directions 
 can also be expressed in the form of (\ref{eq:mubpaulichannel}) on the MUB determined by 
 $|\varphi_{1,1}\>, |\varphi_{2,1}\>,|\varphi_{3,1}\>$ 
 with Bloch vectors $\n{v}_1$, $\n{v}_2$, and $\n{v}_3$. These vectors  
 form a basis in $\mathbb{R}^3$. 
Let us further assume that the input qubit to the Pauli channel is represented by the 
 Bloch vector $\n{b}$ \textit{in the $\{ \n{v}_i \}$ basis representing the channel directions}. 

Then, the effect of the channel for the input Bloch vector $\n{b}$ ($\Vert\n{b}\Vert\leq 1$) can be written as 
\[
\n{b}=\sum_{i=1}^3b_i\n{v}_i\rightarrow\sum_{i=1}^3\lambda_ib_i\n{v}_i\ .
\]

In the rest of this section, the words ``vector'' and ``state'' are used as synonyms, both referring to Bloch vectors. 

\subsubsection{The Case of Different Channel Parameter Values}
Assume that all of the $\lambda_i$ channel parameters have different absolute values, therefore $|\lambda_i|<1~,~i=1,2,3$. 
Prepare a pure state $\tilde{\n{b}}^{(0)}$ with $\big\Vert\tilde{\n{b}}^{(0)}\big\Vert= 1$, and use it as input to the channel. 
Then the output $\n{b}^{(1)}$ can be expanded in the $\{\n{v}_i\}$ basis, and we get that each component of $\tilde{\n{b}}^{(0)}$ got scaled by the corresponding $\lambda_i$ parameter. As $|\lambda_i|< 1$ from the positivity and trace preserving constraints (\ref{eq:trace.preserving}), the absolute value of each nonzero component will get smaller. Let the channel parameter with the largest absolute value be $\lambda_m$. Then the absolute value of component $\tilde{b}^{(0)}_{m}$ will decrease relatively the least among nonzero components of $\tilde{\n{b}}^{(0)}$. In other words, the value $\frac{\tilde{b}^{(n)}_{i}}{\Vert\n{b}^{(n)}\Vert}$ will get bigger for $i=m$, and smaller for $i\neq m$.  

If we continued this procedure, and put the channel output $\n{b}^{(n)}$ back into the channel as input to get the output $\n{b}^{(n+1)}$, then by the above argument, the value $\frac{\tilde{b}^{(n)}_{i}}{\Vert\n{b}^{(n)}\Vert}$ will converge to $1$ for $i=m$, and to $0$ for $i\neq m$. Thus the sequence
\[
\left\{\frac{\n{b}^{(n)}}{\big\Vert\n{b}^{(n)}\big\Vert}\right\}_{n=0}^\infty
\]  
will be a Cauchy sequence, and will converge to the direction $\n{v}_m$ that corresponds to the parameter $\lambda_m$ with the largest absolute value. 
The normalization in the above sequence is inevitable, as the output states do not remain pure during the iterated channel effect, i.e. the length $\big\Vert\n{b}^{(n)}\big\Vert$ of the sequence will not remain $1$, it will converge to zero instead. 

Thus, to prevent the vector sequence from converging to the maximally entangled state, we have to do the normalization of the output Bloch vector $\n{b}^{(n)}$ manually after each step. This means that we have to exchange the output state with the pure state which points in the same direction. Thus we need to perform quantum state tomography. After the normalization of $\n{b}^{(n)}$, we get the pure state $\tilde{\n{b}}^{(n)}$, which can be put again in the channel. This way, the Cauchy sequence of vectors converge to $\n{v}_m$. 

The accuracy of this procedure has of course a limit, set by the accuracy of quantum state tomography. Convergence to a channel direction is guaranteed only until the difference in the input and output state is not comparable with the uncertainty of the state estimation procedure. Thus, when the sequence reaches this limit, the searching procedure should stop. It can also occur that we give a good initial guess, and start with an input state which is close to a channel direction. Then that direction can be accepted, as slow convergence can only occur close to channel directions. 

After the first channel direction $\n{v}_m$ was found using this procedure, we can continue the search in the plane orthogonal to $\n{v}_m$. However, due to the inaccuracies in state tomography, the direction we will find will not be exactly $\n{v}_m$, rather some vector $\tilde{\n{b}}^{*}\approx\n{v}_m$. Thus, it is more robust if we apply a projection to the output vector, onto the subspace in which we want to do the searching. When the second direction is found, then the third can be easily obtained, as it will be the one orthogonal to both the first and the second direction. Thus the direction estimation procedure is finished. 

\subsubsection{The Case of Equal Channel Parameter Values}
In the degenerate cases when some of the channel parameters $\lambda_i$ have equal absolute values, then the channel is equally depolarizing in the linear span of those directions, i.e. there are no exact channel directions defined in that subspace. This means that we can use any state inside this subspace as channel direction, so the sequence $\big\{\tilde{\n{b}}^{(n)}\big\}_{n=0}^\infty$ of states is only required to converge to an arbitrary state inside this subspace, which is guaranteed by the above procedure. 

It follows, that if all the channel parameters have equal absolute values, then the channel is the depolarizing channel, which means that any three orthogonal Bloch vectors can be used as channel directions.


\subsubsection{Algorithm for Direction Estimation}
The procedure described in the previous subsections can be summarized in the following algorithm. We would like to estimate the three depolarizing directions of a qubit Pauli channel $\cha{E}$. Let the set of found channel directions be $\n{D}$. We start with $\n{D}=\{\}$ and $n=0$.
\begin{enumerate}
\item Prepare a pure state $\tilde{\n{b}}^{(n)}\in\n{D}^\perp$. \label{alg:direction.estim.1}
\item Put $\tilde{\n{b}}^{(n)}$ into the composite channel $\cha{E}^k$ formed by cascading $k$ instances of the channel $\cha{E}$, then get the output $\n{b}^{(n+1)}$. \label{alg:direction.estim.2}
\item Perform quantum state tomography on $\n{b}^{(n+1)}$.
\item Normalize $\n{b}^{(n+1)}$ to get the pure state $\tilde{\n{b}}^{(n+1)}$. 
\item Project $\tilde{\n{b}}^{(n+1)}$ to the subspace $\n{D}^\perp$.  
\item If the distance $\big\Vert\tilde{\n{b}}^{(n)}-\tilde{\n{b}}^{(n+1)}\big\Vert$ is smaller than the some value determined by the variance of the used quantum state estimation  method, then continue with step (\ref{alg:direction.estim.7}). Else increase $n$ by $1$ and continue with step (\ref{alg:direction.estim.2}).
\item Put $\tilde{\n{b}}^{(n+1)}$ into the vector set $\n{D}$, set $n$ to $0$ then restart with step (\ref{alg:direction.estim.1}). \label{alg:direction.estim.7}
\end{enumerate}

This algorithm -- though rather resource intensive -- thus estimates the directions of a Pauli channel. During the algorithm, we can get information also on the channel parameters, which can be made more accurate using the optimal tomography configurations described in Section \ref{sec:exp.des}, thus making a two step Pauli channel estimation procedure. 

\subsection{A Simple Numerical Example}

In order to illustrate the operation and properties of the 
 above proposed channel direction estimation algorithm, a simple illustrative 
 numerical example is presented here for a qubit channel with different parameters $\lambda_1=0.6$, $\lambda_2=0.3$ and $\lambda_3=0.1$. 
 
The three unknown channel directions were chosen to be the eigenvectors of the Pauli matrices. 
The uncertainty in the estimated channel output state arising from quantum state tomography was simulated using random perturbations in the output state. The perturbation for the $i$th direction is a random term added to the Bloch vector component $\theta_i$, and is of the form
 \[
 \xi \sqrt{\frac{1-\theta_i}{N}}\ ,
 \]
 where $\xi$ is a random number taken form the standard normal distribution, $N$ is the number of measurements in the state tomography step, and $\frac{1-\theta_i}{N}$ is the variance of the estimator $\hat{\theta_i}$. 
 
\begin{figure}
\begin{center}
\im[width=0.8\textwidth]{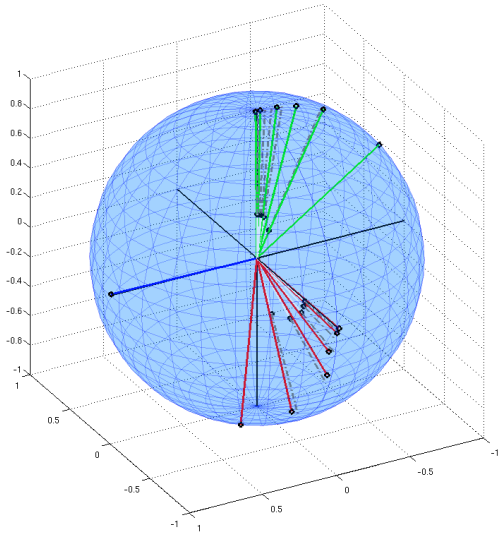}
\end{center}
\caption{Example on channel direction estimation for the qubit channel with parameters $\lambda_1=0.6$, $\lambda_2=0.3$ and $\lambda_3=0.1$. \label{fig:2Ddirest}}
\end{figure}

The result of the numerical test can be seen on Figure \ref{fig:2Ddirest}. 
The three unknown channel directions are shown by the black axes in the Bloch sphere. 
The colored vectors indicate the perturbed and normalized input states in each step, and the white vectors indicate channel outputs. The starting input vector were chosen randomly at the beginning of the search for each direction. The perturbation in the output states assumed $N=5000$ measurements in the state tomography steps. The red vectors correspond to the direction with parameter $\lambda_1$ of the largest absolute value,  the green vectors correspond to the direction with parameter $\lambda_2$, and the blue vector correspond to the remaining direction with parameter $\lambda_3$. Five iteration steps were performed in each of the two estimated 
directions. 

It can be seen from the figure, that the sequence of input states converge to the channel direction of highest absolute parameter value in the subspace $\n{D}^\perp$ in five iteration steps. 
 
\section{Experiment Design in the Known Channel Direction Case}
\label{sec:exp.des}

The field of experiment design for quantum channel parameter estimation 
 has not matured yet, even the problem statements have not cleared up.  
Only a few papers exist that aim at determining the elements of the tomography configuration, i.e.\ the input quantum system and the measurement POVMs, 
(see e.g.\ \cite{Sarovar2006}, \cite{Fujiwara2003}). 
These papers, however, fix one of the elements -- the input quantum system, for example -- 
 and determine the other (say the POVM) according to some optimality criteria. 
The only paper that uses an optimization approach to experiment design 
 solves a  restricted experiment design problem, i.e.\ the 
 determination of the number of measurements to be performed in the different experiment configurations \cite{Kosut-2004}. 
 
In this section an experiment design problem of the whole tomography configuration 
 is formulated and solved in the form of a convex maximization problem. 

\subsection{Problem Statement}
Suppose we have a quantum channel $\cha{E}_{\lambda}$ with some fixed channel parameter vector $\lambda$. We would like to find the input state $\rho$ and a measurement POVM $\n{M}$ for which the Fisher information $F(\lambda)$ of the channel parameters estimated from the channel output $\cha{E}_{\lambda}(\rho)$ using the POVM $\n{M}$ is maximal. As the Fisher information will be a matrix, by maximization of $F(\lambda)$ we mean the maximization of an appropriately selected scalar function of $F(\lambda)$.

As we have seen before, the probability $p(\alpha|\lambda)$ of the measurement outcome $\alpha$ of the state $\cha{E}_{\lambda}(\rho)$ can be rewritten as $\trace{(\rho_\lambda M_\alpha)}=\trace(C_\alpha X_{\lambda})$, where $C_\alpha=(\rho^\tp\otimes M_\alpha)$ is the configuration matrix. Using this, the Fisher information will be 
\begin{equation*}
\label{eq:fisher.info.element-channel}
[F(\lambda)]_{i,j}=\sum_\alpha\frac{1}{\trace(C_\alpha X_{\lambda})}
\frac{\partial}{\partial \lambda_i}\trace\left(C_\alpha X_{\lambda}\right)
\frac{\partial}{\partial \lambda_j}\trace\left(C_\alpha X_{\lambda}\right)\ .
\end{equation*}

To be able to do the maximization, a scalar valued objective function is needed. By the property $\trace(A)\leq\trace(B)$ whenever $A\leq B$ for the Hermitian matrices $A$ and $B$, we take the trace of the Fisher information matrix:
\begin{equation}
\tilde{F}(\lambda)=\sum_{i,\alpha} \frac{1}{\trace(C_\alpha X_{\lambda})}
\left(\frac{\partial}{\partial \lambda_i}\trace\left(C_\alpha X_{\lambda}\right)\right)^2\ .\label{eq:fisher.objective}
\end{equation}

It can be shown that the function $\tilde{F}$ is convex in the configuration matrix $C_\alpha$ on the set of valid $C_\alpha$ matrices, thus convex both in the input $\rho$ and the used measurement POVM $\n{M}$ if we fix the other to be a constant. 


From this it follows that $\tilde{F}$ should take its maximum at an extremal point of the feasible region containing the possible experiment configurations. \textit{Thus, the optimal input state will be pure, and the optimal measurement POVM will be a so called extremal POVM} \cite{Dariano-2008}. 

\subsection{The Optimal Configuration for Qubit Pauli Channels}
Now we study the case of qubit Pauli channels, and will assume that the three depolarizing directions of the Pauli channel are known. Because of the rotational symmetry of the Bloch ball, the obtained results can be applied to any other Pauli channel, with different directions. 
 
The experiment design problem is solved for projective measurements, which can be represented with two-element extremal POVMs $\{\ketbra{\psi}{\psi},I-\ketbra{\psi}{\psi}\}$ \cite{Dariano-2008}. 
Let these POVM elements be represented with the Bloch vectors $\n{m}$ and $-\n{m}$ with $\norm{\n{m}}_2=1$. 
Let also the pure input state be in Bloch parametrization \eqref{E:2x2}, with the Bloch vector denoted as $\n{b}$.

Then the channel output with channel parameter vector $\lambda=[\lambda_1,\lambda_2,\lambda_3]^\tp$ will be (\ref{eq:2x2out}), and if we write the trace of the Fisher information matrix of the the channel parameters, we get
\begin{equation}
\tilde{F}(\lambda)=\frac{m_1^2b_1^2+m_2^2b_2^2+m_3^2b_3^2}{1-(m_1 b_1 \lambda_1 + m_2 b_2 \lambda_2 + m_3 b_3\lambda_3)^2}\ .\label{eq:fisher.qubit}
\end{equation}
Recall that the unit length requirement on the vectors $\n{b}$ and $\n{m}$ follows from the convexity of $\tilde{F}$, which we want to maximize. 
Note also that the above formula is a special case of Eq. (\ref{eq:fisher.objective}). 

Let us now define the vector $\n{c}=[m_1 b_1, m_2 b_2, m_3 b_3]^\tp$, which can be tought of as the configuration vector of the channel estimation problem, which includes not only the input state and measurement information, but in this case also the assumptions on the channel structure. The objective \eqref{eq:fisher.qubit} will then be
\[
\tilde{F}(\lambda)=\frac{\n{c}^\tp \n{c}}{1-(\n{c}^\tp\lambda)^2}=\frac{\n{c}^\tp \n{c}}{1-\n{c}^\tp\lambda\lambda^\tp \n{c}}\ .
\]

By H\"olders inequality it is easy to see that the set of all possible $\n{c}$ vectors form an octahedron inside the Bloch sphere, whose vertices are the unit vectors pointing to the three directions of the channel. Thus the set of all $\n{c}$ vectors is convex, moreover we have equality if and only if $|b_i|^2=|m_i|^2$, i.e. when the vectors $\n{b}$ and $\n{m}$ are parallel. 

As the objective is convex in both $\n{b}$ and $\n{m}$ and thus in $\n{c}$ 
we know that it takes its maximum at a vertex of the octahedral feasible set. Thus the optimal $\n{c}$ has not only unit 1-norm, but unit 2-norm, too. This can only happen if only one component of $\n{c}$ is nonzero, which means that both the input and the measurement have to be in the same channel direction. This implies that the objective is maximized clearly if the direction of $\n{c}$ is that direction, for which $|\lambda_i|$ is maximal. Let this be for example $\lambda_1$, then the optimal objective will be
\[
\tilde{F}(\lambda)=\frac{1}{1-\lambda_1^2}\ .
\]

Now, we see that performing experiments in this direction does not give any information on the other directions, so we have to search for additional experimental configurations. Let the direction of the optimal configuration found first be the direction $x$, i.e.\ for $i=1$. 
If we now constrain the objective \eqref{eq:fisher.qubit} to the plane orthogonal to $x$, then we get the constraints $m_1=0$ and $b_1=0$. Using the same derivation as in the general three dimensional case, we get that the next optimal configuration will be the $y$ (with $i=2$) or $z$ (with $i=3$) direction, and so on. 

As a conclusion, \textit{the three Pauli channel directions can be used as optimal directions for both measurements and input states}. 

\subsubsection{Optimal Parameter Estimation of Qubit Pauli Channels}
The general least squares objective function in Eq. (\ref{eq:ls.objective}) used for process tomography can be simplified using the optimal experiment configuration. 

For the case of qubit Pauli channels, we can express the outcome probabilities in the configuration $\gamma$ as 
\[
p_{\pm,\gamma}=\frac{1\pm\n{c}^\tp\lambda}{2}\ .
\]
Substituting this into the objective function \eqref{eq:ls.objective}, we get
\begin{gather}
\arg\min_{\lambda} \sum_{\pm,\gamma}\left(\hat{p}_{\pm,\gamma}-\frac{1\pm\n{c}^\tp\lambda}{2}\right)^2\ ,\\
\tn{so that}\quad|1\pm\lambda_3|\geq|\lambda_1\pm\lambda_2|\ .\nonumber
\end{gather}
If we assume that the channel is truly Pauli, then we do not need the constraints on the parameters, because the global minimum of the objective function will be inside the feasible region. Applying the optimal tomography configuration, we get 
\begin{gather}
\arg\min_{\lambda} \sum_{\pm,\gamma}\left(\hat{p}_{\pm,\gamma}-\frac{1\pm\lambda_\gamma}{2}\right)^2\ ,
\end{gather}
which can be written as 
\[
\frac{1}{2}\lambda^\tp\lambda+(\hat{\n{p}}_--\hat{\n{p}}_+)^\tp\lambda+\hat{\n{p}}_+^\tp\hat{\n{p}}_++\hat{\n{p}}_-^\tp\hat{\n{p}}_--\frac{3}{2}.
\]
Here the $\hat{\n{p}}_\pm$ vectors contain the measured $+1$ and $-1$ outcome probabilities (i.e.\ the relative frequencies) for each configuration. Setting the gradient equal to zero, we get the optimal estimator for the channel parameters:
\[
\hat{\lambda}=\hat{\n{p}}_+-\hat{\n{p}}_-\ .
\]
Thus if the optimal configurations are used, then the qubit Pauli channels can be estimated in a very simple and efficient way.

\medskip
\subsubsection{Optimal Configuration for Generalized Pauli Channels}
Unfortunately, the above derivation does not generalize to the higher 
 dimensional case in a straightforward way.  
Therefore, numerical optimization can be used to 
 find optimal measurement configurations. A case study can be seen for the qutrit case in subsection \ref{subsec:cs.gen.pauli}. The example suggests that in higher dimensions, when the complementary subalgebras $\mathcal{A}_i$ are maximal Abelian, the optimal set of configurations will be similar to that of two dimensional Pauli channels. 
 
 Namely, we will need $d+1$ configurations, with a pure state $\rho\in\mathcal{A}_i$ as input, and the POVM $\{M_{\alpha,i}\}$ as measurement, where $M_{\alpha,i}\in\mathcal{A}_i$ is also pure in the $i$th configuration. 

\section{Case studies}
\label{sec:case.studies}
The aim of the simulation experiments was to analyze the effect of experiment design on the performance of the numerical optimization based estimation of quantum channels. Results were generated in MATLAB environment, using simulated random measurement data. The optimization problem \eqref{eq:ls.objective} was solved using YALMIP modeling language \cite{YALMIP} and the SDPT3 solver \cite{SDPT3}. 

\subsection{Tomography Configurations}
The experiments in the nonoptimal test cases were set up as follows.
\begin{itemize}
\item The used input states were all pure states. 
\item To obtain a tomographically complete measurement, appropriate POVMs were selected on the Hilbert space of the system, and were used for measurement. In two dimensions, these were the \emph{Pauli matrices}, and observables with similar properties in higher dimensions. Each of these can be decomposed into a POVM, which can be used in one configuration. 
\item The total $n_\tn{tot}$ number of measurements was distributed among all the configurations equally, i.e. for each configuration $\gamma$, an equal number of experiments  were used. 
\end{itemize}
Each experiment setup was repeated five times and their average was taken. Each of the estimated process Choi matrices $\op{X}_\cha{E}$ and channel parameters were analyzed using the following estimation performance measuring quantities:
\begin{itemize}
\item The empirical mean $\bar{\lambda}$ of the estimated parameters $\hat{\lambda}$. 
\item The empirical covariance matrix of the estimated parameters $\hat{\lambda}$, 
where only the diagonal elements of the covariance matrix, i.e. the variances of the parameters were computed as
\[
\tn{Var}(\hat{\lambda}_i)=\frac{1}{4}\sum_{j=1}^5(\hat{\lambda}_{i,j}-\bar{\lambda}_{i})^2\ .
\]
\item Hilbert--Schmidt norm of the estimation error: $\Vert\hat{\op{X}}_{\mathcal{E}}-\op{X}_{\mathcal{E}}\Vert$.
\end{itemize}

\subsection{Qubit Pauli Channels}
\label{ssec:t.repr}
In this section we show some examples on process tomography which demonstrates the differences between nonoptimal and optimal experiment configurations. 
The channel parameters in each test were $\lambda_1=0.3$, $\lambda_2=-0.1$, $\lambda_3=0.1$.

\subsubsection{Pauli Channel Estimation with Nonoptimal Configuration} 
\begin{figure}
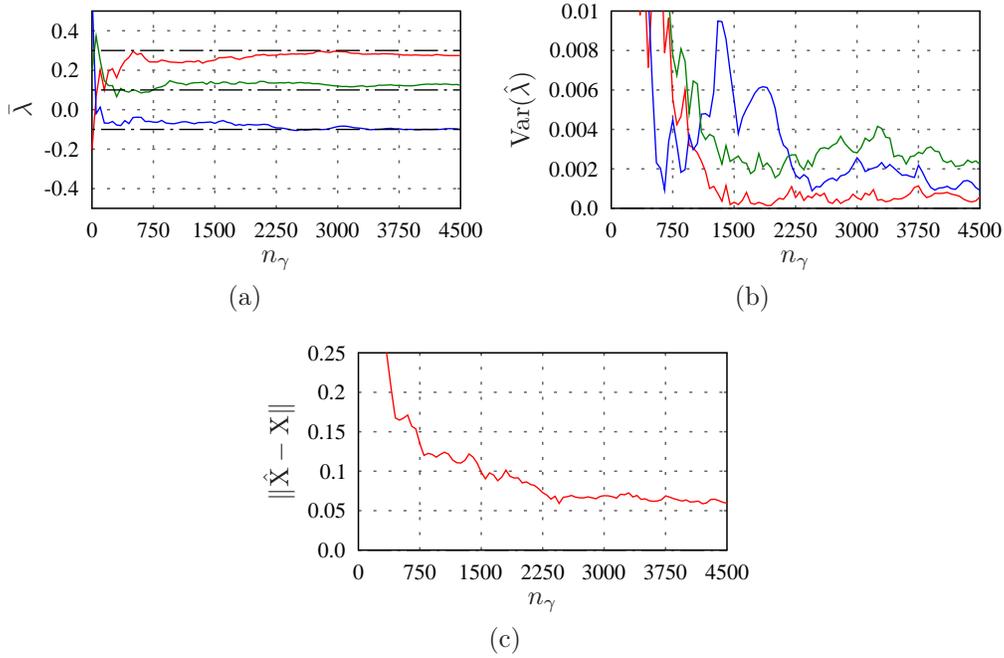

\begin{center}
\subfloat[]
{   
    \im{pau2-min-mea-4500.eps}
    \label{fig:pau2-min-mea}
}
\subfloat[]
{   
    \im{pau2-min-var-4500.eps}
    \label{fig:pau2-min-var}
}\\
\subfloat[]
{   
    \im{pau2-min-hil-4500.eps}
    \label{fig:pau2-min-hil}
}
\end{center}
\caption{Estimation with nonoptimal configuration for the channel parameters $\lambda_1=0.3$, $\lambda_2=-0.1$, $\lambda_3=0.1$. \label{fig:pau2-min}}
\end{figure}

First, we perform an experiment with the minimal POVM described by \cite{Rehacek-2004}, which is tomographically complete. 
In this case only one configuration is used with a total number of measurements
$n_\tn{tot}=n_{\gamma}=4500$, and  with the pure input state
\[
\frac{1}{\sqrt{3}}[1,1,1]^\tp\ .
\] 
The characteristic quantities are plotted in Figure \ref{fig:pau2-min}.

\subsubsection{Pauli Channel Estimation with Nonoptimal Input State} 
\begin{figure}
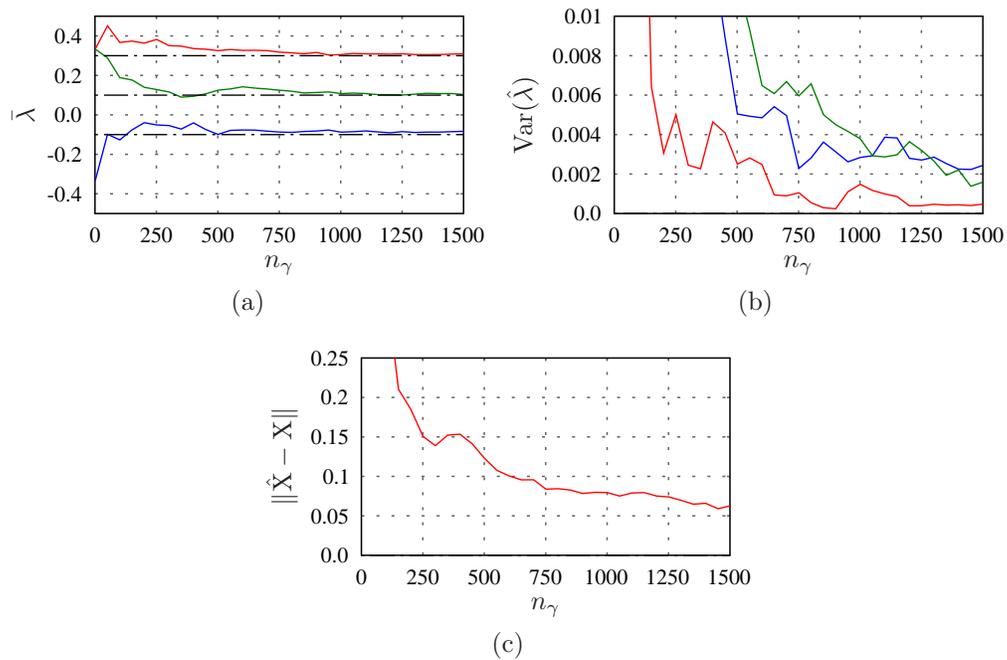

\begin{center}
\subfloat[]
{   
    \im{pau2-std-mea.eps}
    \label{fig:pau2-std-mea}
}
\subfloat[]
{   
    \im{pau2-std-var.eps}
    \label{fig:pau2-std-var}
}\\
\subfloat[]
{   
    \im{pau2-std-hil.eps}
    \label{fig:pau2-std-hil}
}
\end{center}
\caption{Estimation with nonoptimal input state for the channel parameters $\lambda_1=0.3$, $\lambda_2=-0.1$, $\lambda_3=0.1$.\label{fig:pau2-std}}
\end{figure}

In this case, the measurements were performed in the three optimal measurement directions using three experiment configurations, with $n_{\gamma}=1500$ measurements in each. The input state, however, was a nonoptimal pure input state 
\[
\frac{1}{\sqrt{3}}[1,1,1]^\tp
\] 
in all three configurations. The characteristic quantities are plotted in Figure \ref{fig:pau2-std}.

\subsubsection{Pauli Channel Estimation with Optimal Experiment Configuration}
\label{ssec:opt-case-study}
\begin{figure}
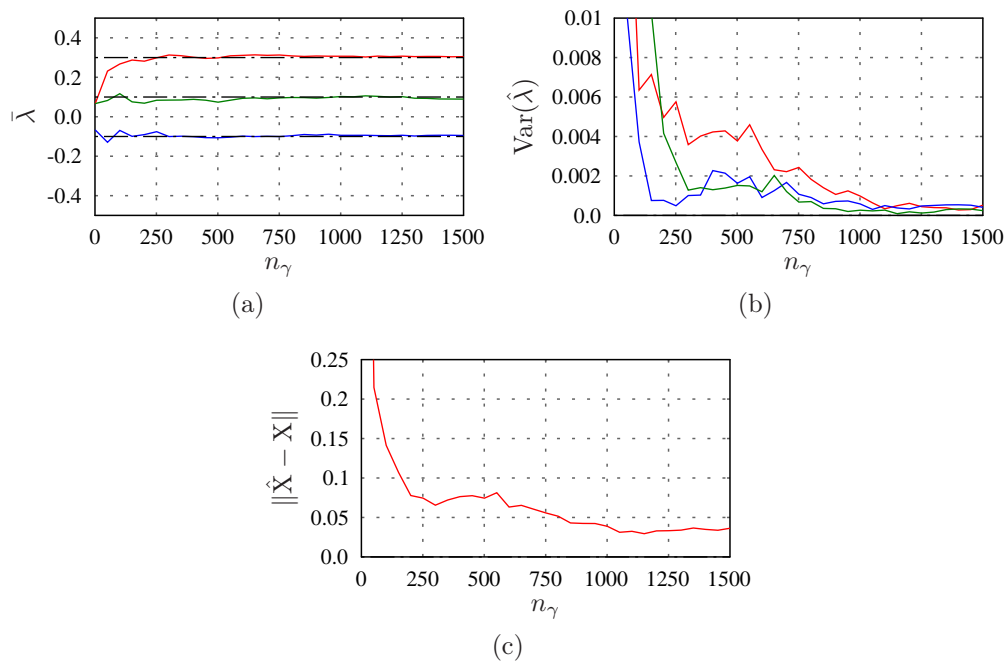

\begin{center}
\subfloat[]
{   
    \im{pau2-opt-mea.eps}
    \label{fig:pau2-opt-mea}
}
\subfloat[]
{   
    \im{pau2-opt-var.eps}
    \label{fig:pau2-opt-var}
}\\
\subfloat[]
{   
    \im{pau2-opt-hil.eps}
    \label{fig:pau2-opt-hil}
}
\end{center}
\caption{Estimation with optimal configuration for the channel parameters $\lambda_1=0.3$, $\lambda_2=-0.1$, $\lambda_3=0.1$.\label{fig:pau2-opt}}
\end{figure}

In this setup, both the input and measurement were optimal with respect to the known channel directions. The characteristic quantities are plotted in Figure \ref{fig:pau2-opt}.

The results indicate, that the efficiency of the optimal 
 experiment configuration highly outperforms the nonoptimal ones. 
We can also see that in the optimal setting, we can reach a very accurate estimation with only about $n_\gamma=1000$ number of measurements in each configuration. 

\subsection{Generalized Pauli Channel in 3 Dimension} 
\label{subsec:gen.pauli}
In this section we show some examples on process tomography which demonstrates the differences between nonoptimal and optimal experiment configurations for the case of the qutrit generalized Pauli channel (i.e.\ with $d=3$), based on numerical studies. The channel parameters in each test were $\lambda_1=-0.3$, $\lambda_2=-0.2$, $\lambda_3=-0.1$ and $\lambda_4=0.1$.

\subsubsection{Nonoptimal Experiment Configuration}
\label{subsec:cs.gen.pauli}
The input state used in this experiment was the pure state
\[
\ket{\Phi}=\frac{1}{\sqrt{6}}\big(\ket{\phi_{1,1}}+\ket{\phi_{2,1}}+\ket{\phi_{3,1}}+\ket{\phi_{4,1}}\big)
\]
which is constructed from the first vector $\ket{\phi_{j,1}}$ of each basis of the MUB, on the analogy of the Bloch vector $\frac{1}{\sqrt{3}}[1,1,1]^\tp$ in the two dimensional case. Based on this analogy, this state is expected to be sufficient for the characterization of the channel parameters. 

As we have discussed before, each channel parameter $\lambda_i$ affects the length of the projected input state $E_i(\rho)$ in the subalgebra $\mathcal{A}_i$ independently. The channel has no other effect. Thus, we can measure the effect of the channel by focusing only on the subalgebras, and estimate the parameters independently. For this purpose, any Hermitian operator belonging to the subalgebra $\mathcal{A}_i$ can be considered as observable, and can be used for the estimation of $\lambda_i$. The reason for this is that the effect of the channel is the same on the whole subalgebra, so it is enough to measure any direction inside $\mathcal{A}_i$ to get information on $\lambda_i$. Moreover, the independence of the measurements on different subalgebras results in a diagonal covariance matrix. 

For example, valid observables for each subalgebra can be constructed using the MUB in the following way: 
\[
\op{A}_i=\sum_{j=1}^dj\ketbra{\phi_{i,j}}{\phi_{i,j}}
\]
This way we get observables with $d$ different eigenvalues. Beyond that, the specific value of the eigenvalues are irrelevant, as we are interested only in the outcome probabilities. 

\begin{figure*}
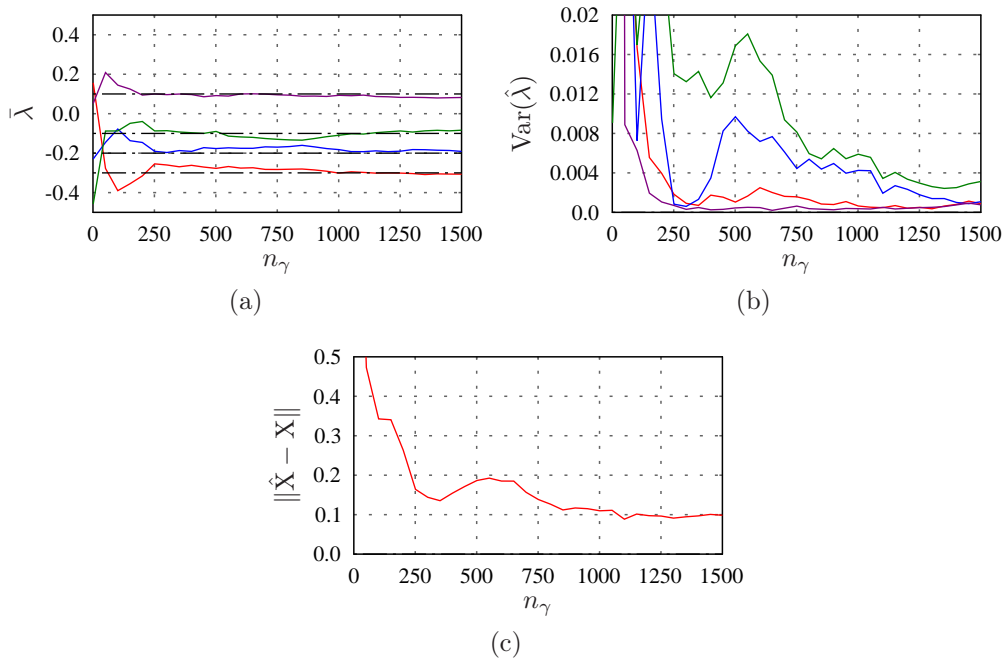

\begin{center}
\subfloat[]
{   
    \im{gpa3-mea.eps}
    \label{fig:gpa3-mea}
}
\subfloat[]
{   
    \im{gpa3-var.eps}
    \label{fig:gpa3-var}
}\\
\subfloat[]
{   
    \im{gpa3-hil.eps}
    \label{fig:gpa3-hil}
}
\end{center}
\caption{Estimation with nonoptimal configuration for the channel parameters $\lambda_1=-0.3$, $\lambda_2=-0.2$, $\lambda_3=-0.1$ and $\lambda_4=0.1$.\label{fig:gpa3}}
\end{figure*}

As it can be seen from the above, a total of four configurations were used in the experiments. The resulting estimations, and characteristic quantities are plotted in Fig. \ref{fig:gpa3}.   More examples can be found in \cite{Ballo-2010}.

We can see that even in the case of a qutrit channel with a high number of parameters, this method can provide very accurate parameter estimation, though the Hilbert--Schmidt norm does not seem to approach zero over the given range of the measurement number $n_\gamma$. Results from experiments performed with higher measurement numbers show that the reason of this is slower convergence. 

\subsubsection{Optimal Experiment Configuration}
In this setup, both the input and measurement were optimal with respect to the known channel directions, as discussed in section \ref{sec:exp.des}. The characteristic quantities are plotted in Figure \ref{fig:pau2-opt}.
\begin{figure}
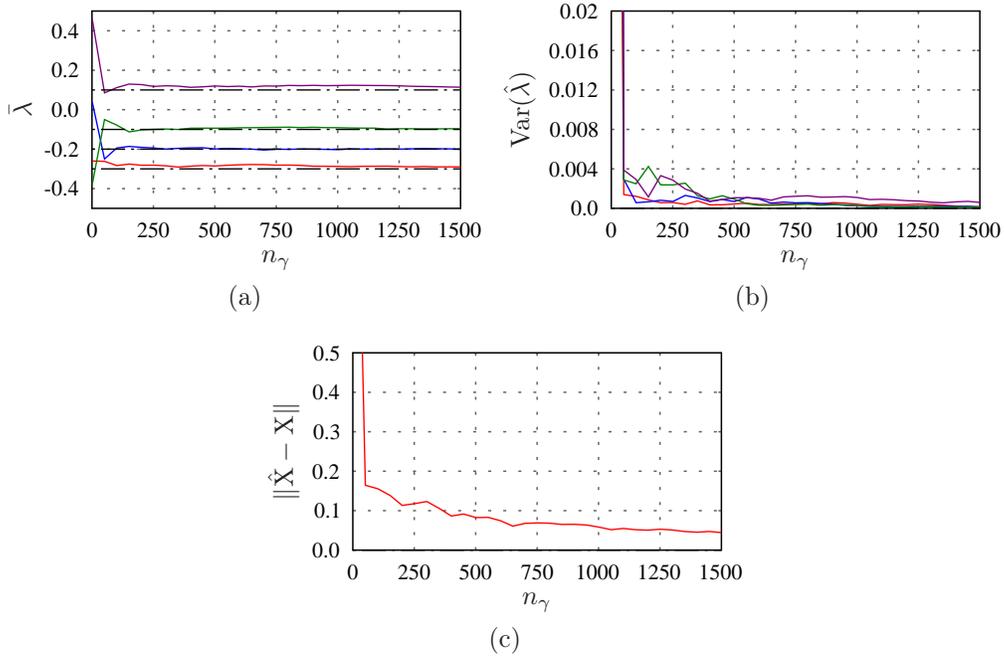

\begin{center}
\subfloat[]
{   
    \im{gpa3-opt-mea.eps}
    \label{fig:gpa3-opt-mea}
}
\subfloat[]
{   
    \im{gpa3-opt-var.eps}
    \label{fig:gpa3-opt-var}
}\\
\subfloat[]
{   
    \im{gpa3-opt-hil.eps}
    \label{fig:gpa3-opt-hil}
}
\end{center}
\caption{Estimation with optimal configuration for $\lambda_1=-0.3$, $\lambda_2=-0.2$, $\lambda_3=-0.1$ and $\lambda_4=0.1$.\label{fig:gpa3-opt}}
\end{figure}

The results indicate, that the efficiency of the optimal experiment configuration highly outperforms the nonoptimal one.  
We can also see that in the optimal setting, we can reach a very accurate estimation with only about $n_\gamma=1000$ number of measurements in each configuration. 

Note that each of the case studies presented in this section assume a known channel model. As we pointed out in section \ref{sec:est.model.families}, knowledge about the channel model could be available in some cases. It can be an interesting question however, that if there is no such information available, then can it be beneficial to try to obtain it, in our case possibly by using the method presented in section \ref{sec:direction_est} to estimate the channel directions? To answer this question, a comparison would be necessary between our method of direction estimation combined with optimal experiment design and a channel estimation method that uses no a priori knowledge about the channel structure, from the aspect of resource requirement. This study is not in the scope of this work, but the papers \cite{Sacchi-2001} and \cite{Kosut-2004} suggest that in order to achieve an accuracy of order comparable with the results given in section \ref{sec:case.studies} without making assumptions on the channel can require a number of measurements of order $10^4$--$10^5$. This is at least about the same order as the approximate measurement requirement of our two step procedure.

\subsection{Robustness of the optimal experiment design}
It is often the case that the optimal experimental configurations designed to probe a set of parameters are sensitive to the assumed parts of the model used to derive the optimal settings. This can be an issue mainly because the direction estimation algorithm presented in section \ref{sec:direction_est} gives only approximate results. Thus, the aim of this subsection is to present a small example on the performance of the optimal experimental configuration for the case when the actual channel directions are slightly perturbed from the assumed channel directions.  

We know, that the channel directions $\ket{\phi_{1,1}}$, $\ket{\phi_{2,1}}$, $\ket{\phi_{3,1}}$ must satisfy \eqref{E:mub}, i.e. each $\ket{\phi_{j,1}}$ must be an element of a basis such that the three bases form a set of MUB. This tells us that the found (possibly inaccurate) channel directions must be transformed versions of the real channel directions by some unitary transform. In the qubit case, expressing the channel directions with a basis of Bloch vectors $\n{v}_1$, $\n{v}_2$, and $\n{v}_3$, this transform can be interpreted as the rotation of each $\n{v}_i$ around a given axis $\n{a}$ with a given angle $\alpha$. 

\begin{figure}
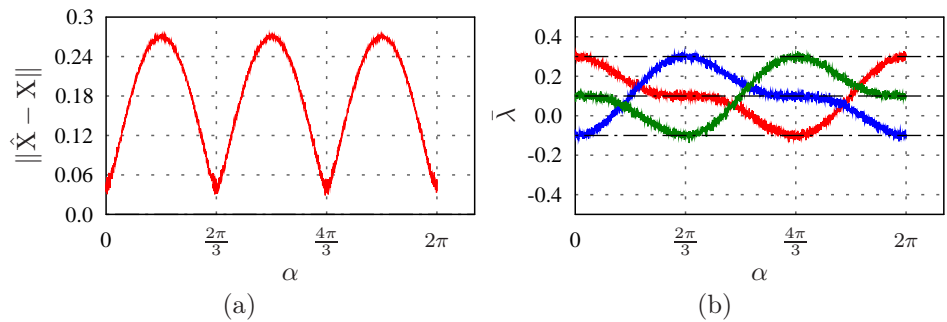

\begin{center}
\subfloat[]
{   
    \im{pau2-robust-hil.eps}
    \label{fig:pau2-robust-hil}
}
\subfloat[]
{   
    \im{pau2-robust-mea.eps}
    \label{fig:pau2-robust-mea}
}
\end{center}
\caption{Example results on the robustness of the optimal experiment design for a qubit Pauli channel with parameters $\lambda_1=0.3$, $\lambda_2=-0.1$, $\lambda_3=0.1$. On Figure \ref{fig:pau2-robust-hil} the Hilbert--Schmidt norm and on Figure \ref{fig:pau2-robust-mea} the estimated parameter means were depicted in function of the perturbation parameter $\alpha$ after $n_\gamma=1500$ measurements in each tomography configuration.\label{fig:pau2-robust}}
\end{figure}

The following example is a modified version of case study \ref{ssec:opt-case-study}. The parameter estimation was done assuming that the $\{\n{v}_i\}$ basis represents the channel directions, but the real channel was simulated using a perturbed basis $\{\n{v}'_i\}$, where $\n{v}'_i=R_\n{a}(\alpha)\n{v}_i$, the matrix $R_\n{a}(\alpha)$ being a rotation matrix. The axis of rotation $\n{a}$ was given by the Bloch vector $\frac{1}{\sqrt{3}}[1,1,1]^\tp$. On Figure \ref{fig:pau2-robust-hil} the Hilbert--Schmidt norm and on Figure \ref{fig:pau2-robust-mea} the estimated parameter means were depicted in function of $\alpha$ after $n_\gamma=1500$ measurements in each tomography configuration. The Hilbert--Schmidt norm is clearly periodic. This is because in this example, after rotating the $\{\n{v}_i\}$ basis by the angle $\alpha=\frac{2\pi}{3}$, we get the basis which is nothing but the starting basis $\{\n{v}_i\}$ with the order of the basis vectors permuted. This can be also seen on Figure \ref{fig:pau2-robust-mea}, where we get the valid parameter values again after rotating by $\alpha=\frac{2\pi}{3}$, just in a different order. Of course from the aspect of robustness, we are only interested in small perturbations, i.e. small values of $\alpha$. 

From this example, it can be seen, that considering the Hilbert--Schmidt norm, the optimal experimental configurations are indeed sensitive to the accuracy of the assumed channel directions, as the norm changes linearly for small perturbations with the perturbation parameter $\alpha$ (see Figure \ref{fig:pau2-robust-hil}). However, the results on Figure \ref{fig:pau2-robust-mea} suggest that the estimated mean values of the parameters change only quadratically for small $\alpha$ values. This means that the optimal parameter estimation method is robust in this sense. 

\section{Conclusions} 
\label{sec:conclusion}
Convex optimization-based parameter estimation and convex maximization-based experiment design methods were 
proposed in this paper for Pauli channels and for their generalized versions. 

A novel method for the parameter estimation of Pauli channel model families were 
 developed for the known channel direction case based on convex optimization.
This method results in a purely convex optimization problem, thus we can obtain a globally optimal estimation with relatively simple and efficient numerical algorithms even in the generalized Pauli channel case. 

Furthermore, an efficient iterative method of estimating the channel directions was also proposed 
 for the qubit Pauli channel case. 
The extension of this method to the general higher dimensional case 
 is a possible direction of our further work.  
 
An experiment design procedure based on maximizing the Fisher information of the output of a generalized Pauli channel is also presented here. It is shown that the Fisher information is a convex function both in the input and in the measurement parameters. Therefore the optimal input state should be pure and the measurement POVM should be extremal. 
For qubit Pauli channels this formulation leads to an optimal setting that includes pure input states and projective measurements directed towards the channel directions. A simple way of estimating the channel parameters in the optimal configuration is also given, and the robustness of the optimal configuration was considered. 
Further work will be directed towards the analytical generalization of this result to the 
 generalized Pauli channel case. 

The effect of the optimal configuration compared to other widely used ones on the parameter estimation performance is demonstrated using case studies. 

\section*{Acknowledgement}
The authors would like to thank L\'{a}szl\'{o} Ruppert for helpful discussions, and advices regarding the manuscript. 

This research was supported in part by the Hungarian Research Fund through grant K67625.

\bibliographystyle{diploma}
\bibliography{Quantum}

\end{document}